\documentclass[sort&compress,
    ,final            
  ]
  {aipproc}

\layoutstyle{6x9}
\usepackage{graphicx}
\usepackage{epsfig}

\begin{document}
\title{Event-based simulation of interference with alternatingly blocked particle sources\footnote{FPP6 -
Foundations of Probability and Physics 6, edited by A. Khrennikov et al., (AIP Conference Proceedings,
Melville and New York, in press)}}


\author{K. Michielsen}{%
  address={Institute for Advanced Simulation, J\"ulich Supercomputing Centre, \\
           Research Centre J\"ulich, D-52425 J\"ulich, Germany},
}
\author{S. Mohanty}{%
  address={Institute for Advanced Simulation, J\"ulich Supercomputing Centre, \\
           Research Centre J\"ulich, D-52425 J\"ulich, Germany},
}

\author{L. Arnold}{%
  address={Institute for Advanced Simulation, J\"ulich Supercomputing Centre, \\
           Research Centre J\"ulich, D-52425 J\"ulich, Germany},
}

\author{H. De Raedt}{%
 address={Department of Applied Physics, Zernike Institute for Advanced Materials,\\
          University of Groningen, Nijenborgh 4, NL-9747 AG Groningen, The Netherlands},
}

\begin{abstract}
We analyze the predictions of an event-based corpuscular model for interference
in the case of two-beam interference experiments in which
the two sources are alternatingly blocked.
We show that such
experiments may be used to test specific predictions of the corpuscular model.
\end{abstract}

\maketitle

\newcommand\sumprime{\mathop{{\sum}'}}
\newcommand{\onlinecite}{\cite}
\def\Eq#1{(\ref{#1})}
\section{Introduction}
Two slit interference experiments have been used to probe fundamental
properties of light and matter for over two centuries. While
Young's original experiment helped establish the wave theory of light~\cite{YOUNG}, variants
of the experiment over the years have helped revealing fascinating wavelike
properties of material entities from electrons~\cite{MERL76,TONO89} to the comparatively humongous
bucky balls~\cite{arndt_wave-particle_1999}.

One interesting variant involves a very dim source so that on average only
one photon is emitted by the source at any time. Photons are also detected one at
a time, as distinctly unwavelike ``clicks`` at the detector, at seemingly
random positions~\cite{JACQ05}. Yet, when a large number of these clicks have been collected,
their locations collectively form precisely the same kind of interference
pattern as would be expected from a wave interacting with the slits.
Such experiments have helped the development of ideas on concepts such as
wave-particle duality in quantum mechanics.

And yet, interference in itself does not entail that the participants be waves.
So long as a concept of an oscillating phase can be associated with individual
particles, even a particulate model can produce interference. Feynman's path
integral description is an example~\cite{feynman1985qed}.
In this paper, we present a toy model for
photon interference effects which are {\it created by} the detector.
This is in contrast to the standard picture where each particle interferes
with itself and the detector passively detects the result. The inclusive results
(averages) from the model about interference are very similar to those obtained
from quantum mechanics, even though each particle has a definite trajectory
through one or the other slit to the detector. We also discuss more
detailed experiments that may be able to distinguish between predictions of this
model and standard calculations.

\section{Model and methods}
In our model~\cite{JIN10b,MICH11a}, which we will refer to as the
''Event Based Corpuscular Model``
(EBCM), each photon carries a 2D vector $\vec{e}$~which oscillates in time with
the photon's frequency according to
\begin{equation}
\label{eq:vece}
\vec{e}=(\cos \omega t, \sin \omega t)
.
\end{equation}
The two slits are modelled as a pair of coherent sources
stochastically  emitting photons isotropically. We imagine that the detector
screen consists of an array of detector elements (pixels) which react
to the incoming photons in the following way. The detector element has a
vector property (e.g. polarisation) which responds to the photon's vector $\vec{e}$ as:
\begin{equation}
\label{eq:memory}
\vec{p}_i = \gamma \vec{p}_{i-1} + (1-\gamma) \vec{e}_i, ~~ 0<\gamma<1
.
\end{equation}
In other words, $\vec{p}_i$ responds to the incoming photon vector $\vec{e}_i$,
but keeps a memory of its previous state. The parameter $\gamma$ controls how
slowly the detector reacts to one individual incident photon.
We assume that the detector element ''clicks`` when the length of the vector $\vec{p}$
reaches a certain threshold.

Simulated photons arrive at one of the detector elements and change its state
according to Eq.~(\ref{eq:memory}). Every once in a while, the vector length at a
detector element reaches the click threshold and an event is recorded. We then
perform a simulation and collect statistics of the click counts along the
detector elements~\cite{JIN10b,MICH11a}.

Several sets of simulations are performed to study the effect of the memory
parameter $\gamma$ and another parameter $N$. The parameter $N$ is used to examine
in detail the effects of prior knowledge of the path of each photon. In the
simulations each source is allowed to emit $N$ photons while the other source is
blocked, and then the states of the two sources (blocked/unblocked) is
interchanged.

\section{Results}
\begin{figure}[t]
\label{setup}
\hspace{10pt}
\includegraphics[width=6cm]{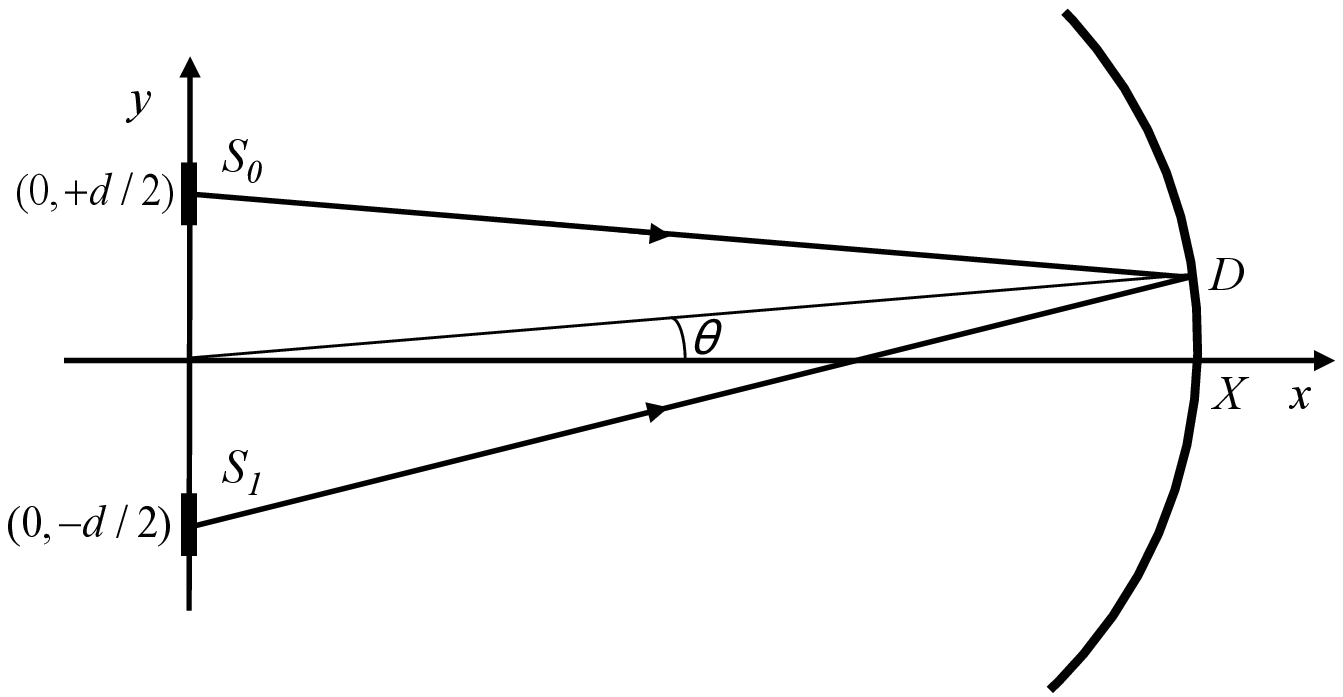}
\includegraphics[width=9cm]{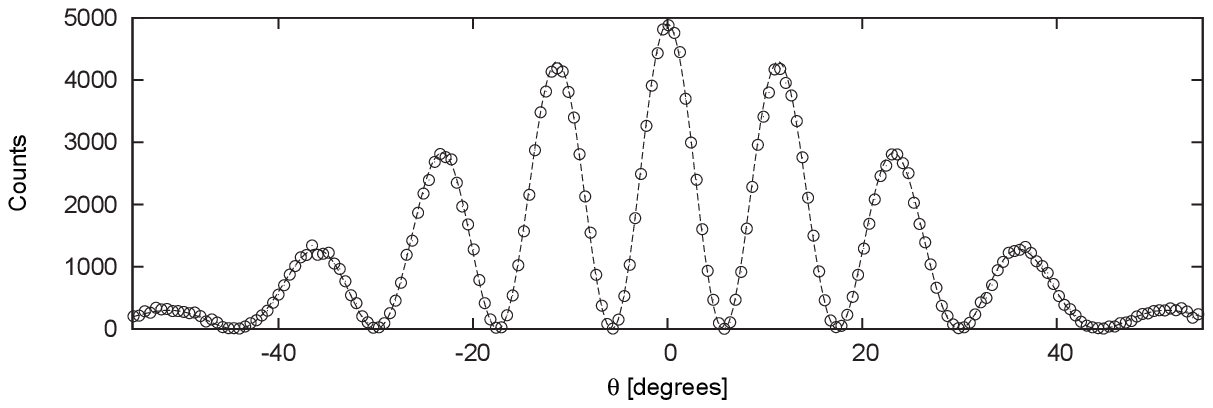}
\caption{a) Basic setup for all comparisons. The distance between the slits
$S_1$ and $S_2$~is $d=5\lambda$, while the slits have a width $a=\lambda$ with $\lambda=670$nm. The circular detector
surface is assumed to be at a large distance $X=0.05$ mm from the sources. (b) Comparison of
general interference results from EBCM simulations (circles) and the regular
optics calculation (solid line).}
\end{figure}

Fig.~\ref{setup}(a) shows the two slit setup for which we show our
results, and compare with the standard optics calculation. The intensity
$I(\theta)$ observed at an angle $\theta$ far away from the two slits is given
by
\begin{equation}
\label{eq:standardtwoslit}
I(\theta)=A\left(\frac{\sin(a\pi\sin\theta/\lambda)}{a\pi\sin\theta/\lambda}
\right)^2 \cos^2(\pi d \sin\theta/\lambda)
,
\end{equation}
where $d$ is the separation between the centres of the slits, $a$ the width of
the slits and $\lambda$ the wavelength.

For a basic test, we compare in Fig.~\ref{setup}(b) the interference result
given by Eq.~(\ref{eq:standardtwoslit})  and the one obtained from simulations with
EBCM. For each event in simulations with EBCM, an emission point
along the lengths of one of the two slits, as well as a direction of travel for
the photon are chosen at random (uniformly distributed). Like in experiments with low intensity,
detector pixels in simulations record events apparently at random locations, but
the interference pattern in Fig.~\ref{setup}(b) builds up over time, in
complete agreement with Eq.~(\ref{eq:standardtwoslit}). The EBCM therefore passes the
first basic test as a model for interference with individual particles.

In standard two slit interference calculations using quantum mechanics, the
wave function of the particles interacts with the two slits, and produces
interference in the amplitude at the detector. The probability of a particle
arriving at a certain location on the detector then turns out to be just like in
classical optics. The particles do not produce interference patterns due to
their mutual interactions, but rather the wave function of individual particles
interferes with itself after being filtered by the two slits. Only after a large
number of events have been measured can the probabilities be compared with the
calculations. On the surface, the EBCM behaves similarly. However, one can think
of more detailed experiments to distinguish it from standard quantum mechanical
calculations.

In particular, in EBCM, we can have knowledge of which slit the photons travel
through, and yet have interference. In quantum mechanics, the wave function
needs to have a non-zero amplitude at the location of both the slits in order
to produce interference. Experiments can be designed for which the predictions
from EBCM differ from quantum mechanics. Consider a variant of the two slit
experiment in Fig.~\ref{setup}, in which the slits open and close periodically
so that at any time there is only one slit open. For simplicity, one can model
the situation like this: the sources $S_1$ and $S_2$ alternatingly emit groups
of $N$ photons each, until a total of $M$ photons have been emitted.

\begin{figure}[ht]
\label{fig:varyN}
\begin{tabular}[t]{ll}
 \includegraphics[width=4.5cm]{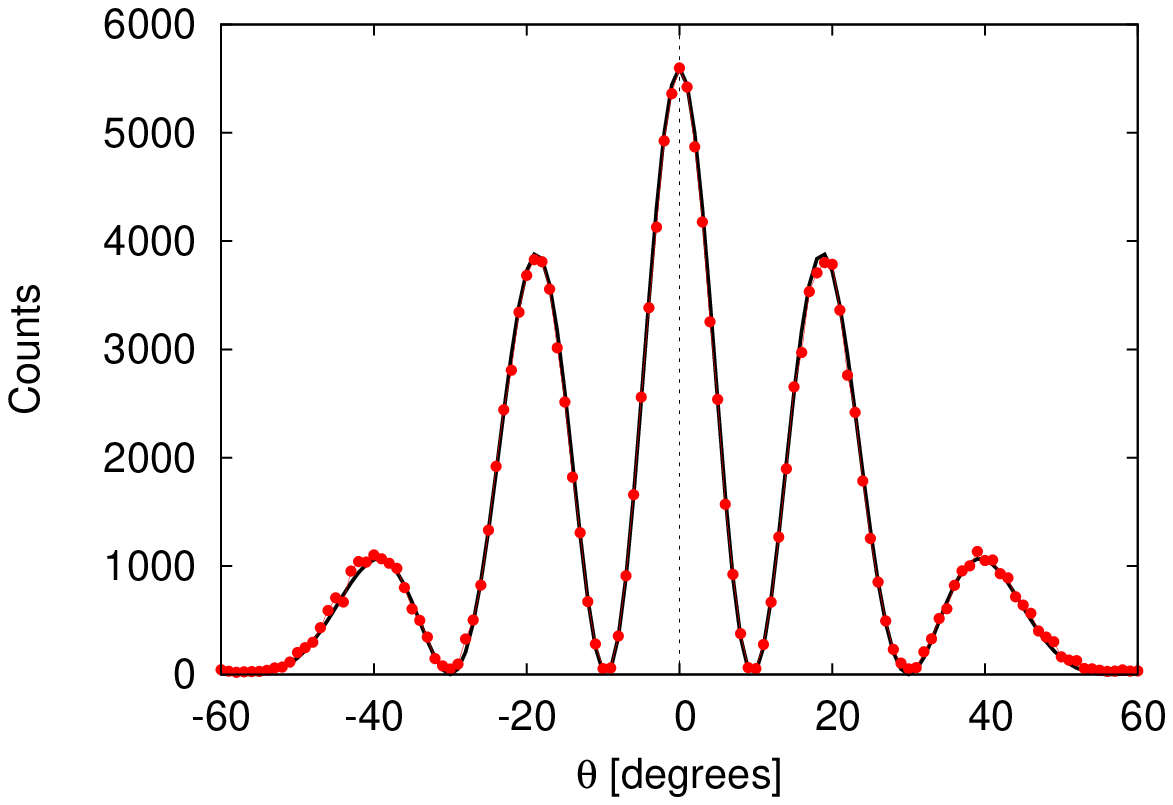} &
 \includegraphics[width=4.5cm]{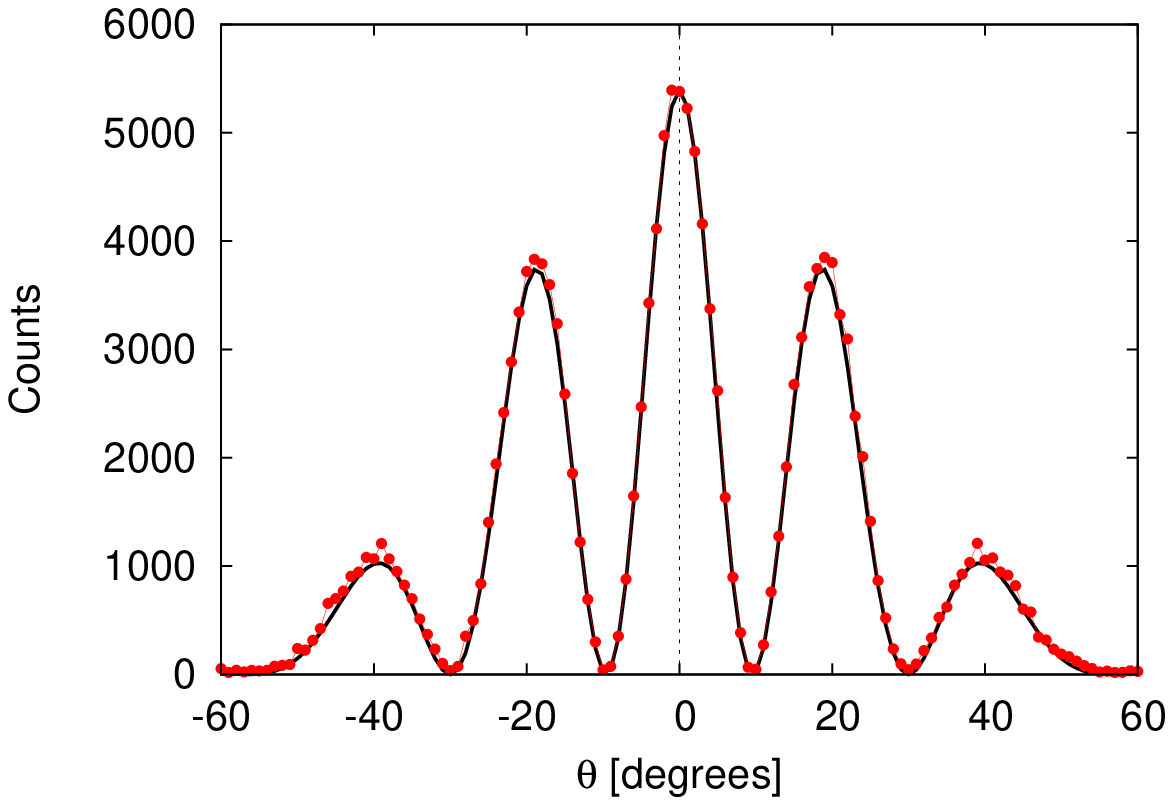} \\
 \includegraphics[width=4.5cm]{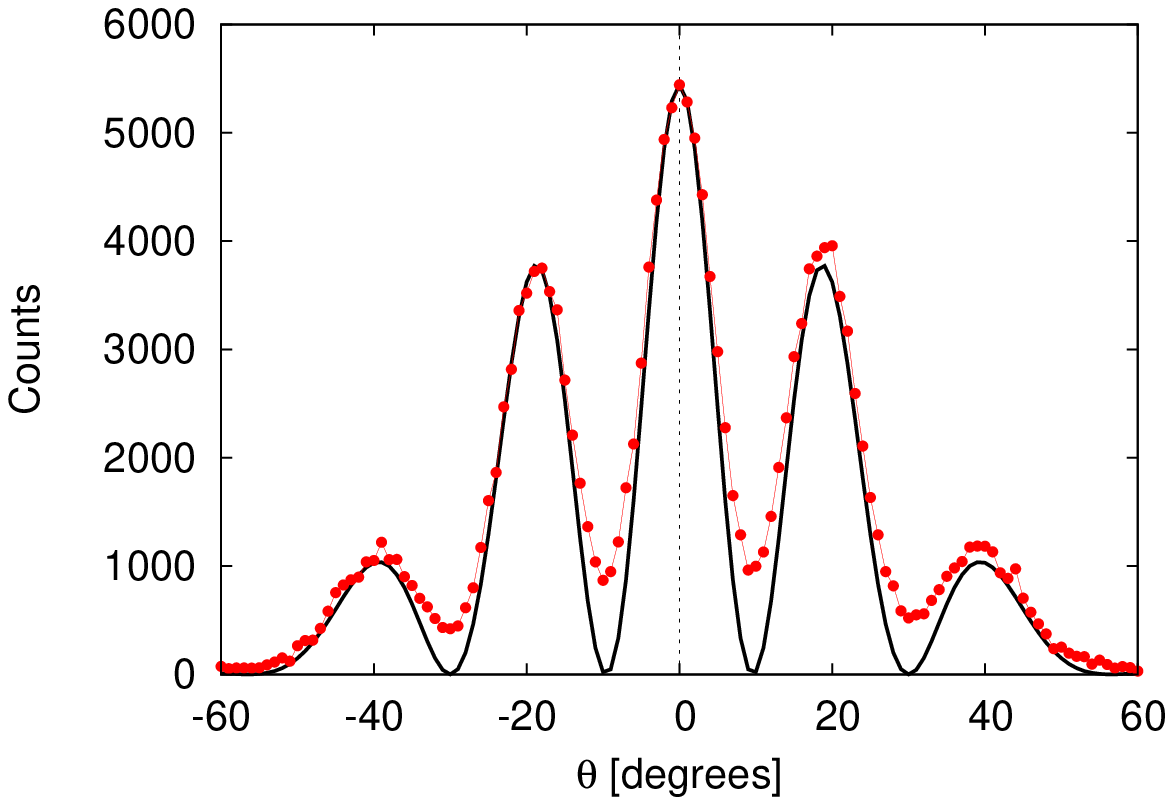} &
 \includegraphics[width=4.5cm]{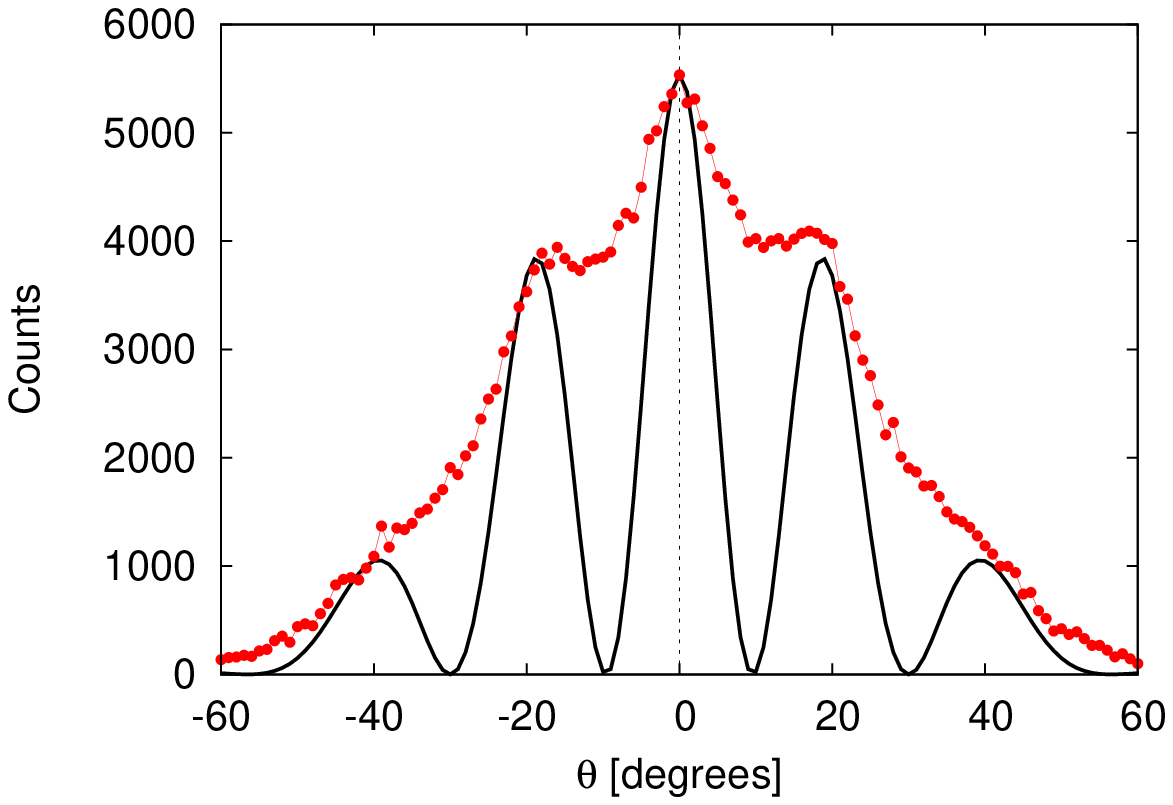} \\
\end{tabular}
\caption{EBCM results for slit switching experiments. Each source $S_1$ and
$S_2$ in Fig.~\ref{setup} emits $N$ photons after which it is switched
off and the other source is switched on. From top left to bottom right (row
first) we show the EBCM simulation results (red dots) for $N=1$, $N=1000$,
$N=100000$ and $N=500000$. For all plots here, $\gamma=0.999$, and the total
number of emitted photons $M=1000000$. The solid line represents the intensity given in Eq.~(\ref{eq:memory})}
\end{figure}

In Fig.~\ref{fig:varyN}, we show EBCM results for slit switching experiments of
different $N$. Evidently, except at very large values of $N$, the interference
pattern remains unchanged. In the simplest calculation with quantum mechanics,
the intensity pattern at the detector for this situation should be the sum of
two single slit patterns, as the particles pass through one or the other slit,
and never through both.

\begin{figure}[ht]
\label{fig:varyg}
\begin{tabular}[t]{ll}
 \includegraphics[width=4.5cm]{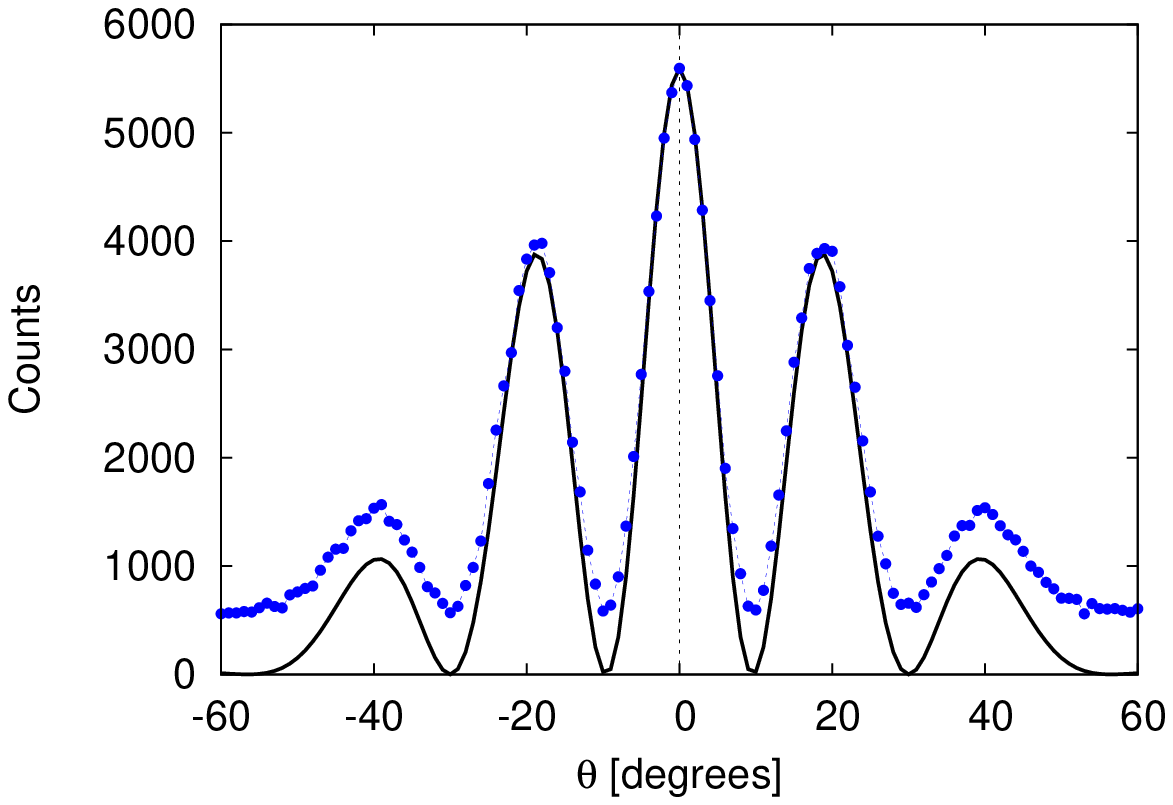}
 \includegraphics[width=4.5cm]{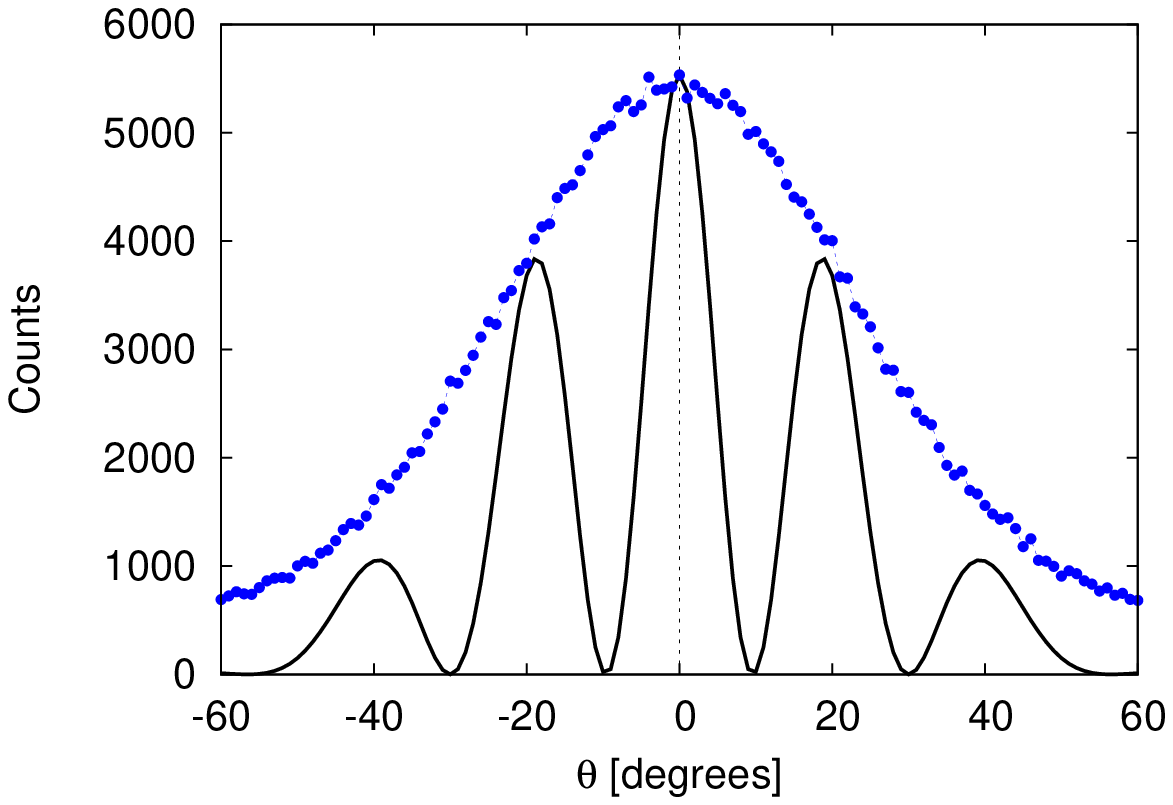}
\end{tabular}
\caption{Same as Fig.~\ref{fig:varyN} for $\gamma=0.9$. Shorter memory,
i.e., smaller $\gamma$ leads to a smoothening of the interference pattern in
EBCM compared to the results shown in Fig.~\ref{fig:varyN}. Left: $N=1$. Right:
$N=500000$. }
\end{figure}

The interference patterns seen in EBCM are a result of memory effects in the
detection units of the experimental setup. This is parametrised by $\gamma$ in Eq. \ref{eq:memory}. In
Fig.~\ref{fig:varyg}, we show EBCM results for a smaller $\gamma=0.9$ compared
to Fig.~\ref{fig:varyN}. By comparing the figures for the corresponding $N$
values, we see that smaller $\gamma$ value smooths out the interference
pattern in EBCM.

\section{Discussion}
In its present formulation, the event based corpuscular model discussed in this
paper, is not a general physical concept with which to understand the physical
world. It might not be applicable to electrons, as the
arrival of the electrons at the detector can be measured  by the electric charge
they carry to the individual detector elements.
In that case, the detector elements react to each individual incident electron with a
``click'' and the threshold behaviour of the modelled detection elements becomes
superfluous. Hence, application of the EBCM discussed in this paper to a similar
experiment with electrons would produce no interference at all. Even in the case
of photons, the ECBM is not an attempt to tell us what the
photons ''actually do``. The EBCM is just a convenient model to describe the
observations in standard two slit experiments.

Nevertheless, sticking to the model idea that interference arises due to state
memory in the experimental setup, there may be measurable consequences that are
different from quantum mechanics. One kind of such experiments would be those
in which only one of the two slits is available to the particles at a time. In
EBCM the results depend on the switching frequency in relation to the photon
emission rate.

Although we are not aware of any experiment that precisely tests the above
scenario, one experimental study in which only one slit was available to each
photon~\cite{daniel_l._aoeeither-ora_2001} produced intriguing results. In that
study, an opaque barrier, all the way from the laser source to the obstacle
between the two slits, was used to make sure that photons had one or the other
slit available to them. The interference pattern observed was nevertheless
essentially unchanged despite the barrier. We are, however, not aware of any
follow-up work on that study.

As a final note, we observe that the model presented here is different from the
''particles with clocks`` picture in Feynman's delightful explanation of the
path integral description of light propagation~\cite{feynman1985qed}. The path
integral formulation is essentially quantum mechanical: the amplitude was
obtained by summing over all possible paths. It does not depend on detector
memory. In the switching slits experiments proposed above, it leads to the
same predictions as quantum mechanics.

\section{Summary}
We have discussed an event based corpuscular model for interference of photons
in a two slit experiment. In this model, interference arises due to memory
effects in the experimental setup. For the most basic setup of the two slit
experiment, the model has identical predictions compared to quantum mechanics.
But it differs from quantum mechanics in predictions about modified experiments
in which only one of the two slits is available to the particles at any time,
and predicts a slow degradation of the interference signal as the switching
between the frames is made progressively slower in relation to the rate of
photon emission. More careful calculations using quantum mechanics as well as
new experiments using today's technology could clarify the picture.

\bibliographystyle{aipproc}       
\bibliography{twoslit,/d/papers/epr11}   

\begin{thebibliography}{9}
\expandafter\ifx\csname natexlab\endcsname\relax\def\natexlab#1{#1}\fi
\providecommand{\enquote}[1]{``#1''}
\expandafter\ifx\csname url\endcsname\relax
  \def\url#1{\texttt{#1}}\fi
\expandafter\ifx\csname urlprefix\endcsname\relax\def\urlprefix{URL }\fi
\providecommand{\eprint}[2][]{\url{#2}}

\bibitem[Young(1802)]{YOUNG}
T.~Young, \emph{Philos. Trans. R. Soc. London} \textbf{92}, 12--48 (1802).

\bibitem[Merli et~al.(1976)]{MERL76}
P.~G. Merli, G.~F. Missiroli, and G.~Pozzi, \emph{Am. J. Phys.} \textbf{44},
  306--307 (1976).

\bibitem[Tonomura et~al.(1989)]{TONO89}
A.~Tonomura, J.~Endo, T.~Matsuda, T.~Kawasaki, and H.~Ezawa, \emph{Am. J.
  Phys.} \textbf{57}, 117--120 (1989).

\bibitem[Arndt et~al.(1999)]{arndt_wave-particle_1999}
M.~Arndt, O.~Nairz, J.~{Vos-Andreae}, C.~Keller, G.~van~der Zouw, and
  A.~Zeilinger, \emph{Nature} \textbf{401}, 680--682 (1999).

\bibitem[Jacques et~al.(2005)]{JACQ05}
V.~Jacques, E.~Wu, T.~Toury, F.~Treussart, A.~Aspect, P.~Grangier, and J.-F.
  Roch, \emph{Eur. Phys. J. D} \textbf{35}, 561--565 (2005).

\bibitem[Feynman(1985)]{feynman1985qed}
R.~Feynman, \emph{QED: The strange theory of light and matter}, Universities
  Press, 1985.

\bibitem[{Jin} et~al.(2010)]{JIN10b}
F.~{Jin}, S.~{Yuan}, H.~{De Raedt}, K.~{Michielsen}, and S.~Miyashita, \emph{J.
  Phys. Soc. Jpn.} \textbf{79}, 074401 (2010).

\bibitem[{Michielsen} et~al.(2011)]{MICH11a}
K.~{Michielsen}, F.~Jin, and H.~{De Raedt}, \emph{J. Comp. Theor. Nanosci.}
  \textbf{8}, 1052 -- 1080 (2011).

\bibitem[Alkon(2001)]{daniel_l._aoeeither-ora_2001}
D.~L. Alkon, \emph{Biophysical Journal} \textbf{80}, 2056--2061 (2001).

\end{thebibliography}

\end{document}